# World Shares of Publications of the USA, EU-27, and China Compared and Predicted Using the New Interface of the *Web-of-Science* versus *Scopus*

Loet Leydesdorff

Amsterdam School of Communications Research (ASCoR), University of Amsterdam, Kloveniersburgwal 48, 1012 CX Amsterdam, The Netherlands;

loet@leydesdorff.net; http://www.leydesdorff.net.

**Abstract**

The new interface of the *Web of Science* (of Thomson Reuters) enables users to retrieve sets larger than 100,000 documents in a single search. This makes it possible to compare publication trends for China, the USA, EU-27, and smaller countries, with the data in the *Scopus* database of Elsevier. China no longer grew exponentially during the 2000s, but linearly. Contrary to previous predictions on the basis of exponential growth, the cross-over of the lines for China and the USA is postponed to the next decade (after 2020) according to this data. These long extrapolations, however, should be used only as indicators and not as predictions. Uncertainty in trends can be specified using the coefficient of determination of the regression ($R^2$) and confidence intervals. Along with the dynamics in the publication trends, one also has to take into account the dynamics of the databases used for the measurement.

## 1. Introduction

On March 28, 2011, the BBC-online had a headline that the Royal Society—the UK's national science academy—had issued a report warning that "China (was) 'to overtake US on science' in two years" based on Elsevier's *Scopus* data (Clarke *et al.*, 2011; Plume, 2011; see Figure 1).

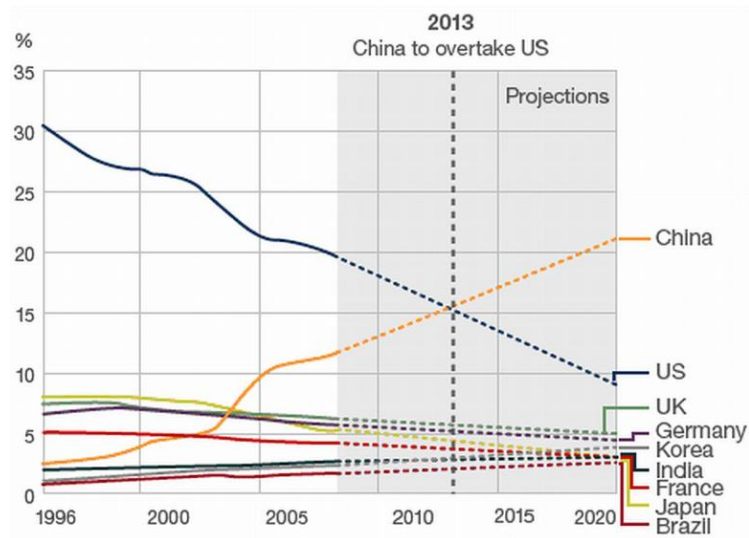

**Figure 1**: Linear extrapolation of future publication trends based on *Scopus* (1996-2008); Source: Clarke *et al*, 2011, Figure 1.6, at p. 43.[1]

In the weeks thereafter, this news led to discussions on the email listing of the US National Science Foundation's "Science of Science Policy" listserver (at scisip@listserv.nsf.gov) about the quality of the prediction based on *Scopus* data. More recently, that is, in July 2011, Thomson Reuters launched Version 5 of the Web-of-Science (WoS) which allows the user—as in *Scopus*—to search directly for countries' shares of contributions, whereas in the previous version one had to overcome indirectly the limits of a recall of more than 100,000 publications in each search (Arencibia-Jorge *et al.*, 2009).

---

[1] We use the remake of the figure by the BBC at http://www.bbc.co.uk/news/science-environment-12885271.

Both *Scopus* and the *Science Citation Index* now allow for direct access to large numbers in the retrieval. In this communication, the new WoS-version of the *Science Citation Index-Expanded* (SCIE) is first used to show the long-term trends of a few leading nations in science and also some smaller ones. The ten-year trendlines for the USA, China, and the EU-27 can be compared using confidence intervals (at the 95% level) for the prediction. These results are compared with those of the Royal Society and the latter will be reproduced using the online version of *Scopus*, but including data for 2009 and 2010. However, the team of Elsevier and the Royal Society, used *Scopus* including the social sciences and humanities, while these were not included using *SCIE* for the measurement. After correction for this, the decline of both the EU-27 and the US since 2004 disappears using Scopus data. The significant differences between using the two databases and different assumptions for the measurement raise questions about the reliability of the prediction.

**2. Theoretical relevance**

The measurement of national publication outputs has been a methodological issue on the research agenda of scientometrics from the very beginning of the *Science Citation Index*. Both Narin (1976) and Small & Garfield (1985) conceptualized this database as a matrix organized along the two dimensions of journals versus countries. The "decline of British Science" in the 1980s (under the Thatcher government), for example, spurred a debate about whether such a decline could perhaps be a scientometric artifact of parameter choices (Anderson et al., 1988; Braun *et al.*, 1989 and 1991; Leydesdorff, 1988 and 1991; Martin, 1991).

At the time, the main database used for the *Science (and Engineering) Indicators* of the US National Science Board (since 1982)[2] was based on two assumptions made by the contracting firm (at the time, Narin's Computer Horizons Inc.): (1) internationally coauthored articles were attributed proportionally to the contributing nations (this is also called "fractional counting") and (2) a fixed journal set was extracted from the *Science Citation Index* for the purpose of longitudinal comparisons (Narin, 1986). Leydesdorff (1988) argued that both these assumptions had an effect on the measurement of output of nations: the ongoing internationalization of coauthorship patterns decreased the national output *ceteris paribus*, and authors in advanced nations such as the UK can be expected to publish above average in new journals associated with newly developing fields of science.

The issue led to a debate and eventually a special issue of *Scientometrics* in 1991 (Braun *et al.*, 1991). Braun *et al.* (1989) distinguished 28 possible parameter choices. The sensitivity of the measurement for relatively minor decisions at the methodological level questions the role of policy advice based on these trendlines for both nations and units at lower levels of aggregation (Leydesdorff, 1989; 1991). How reliable is this data for comparisons among years? One would expect random fluctuations to be averaged out at a high level of aggregation, and thus uncertainty to be reduced. Nowadays, one can additionally ask whether the two major databases (*Scopus* and the WoS) can provide us with similar results. What may be sources of misspecification and therefore potential misrepresentations in the policy arena (Leydesdorff, 2008)?

---

[2] The 1988 edition of these indicators was named "Science & Engineering Indicators". Before this date the title was "Science Indicators".

The issue of the competion of China as a leading nation in science is particularly salient to the science-policy debate today. How much of the spectacular increase of the Chinese world share of publications during the 1990s and 2000s can be attributed to internationalization which goes to the detriment of national publication outlets (Wagner, 2011)? Zhou & Leydesdorff (2006) conjectured that different from linear growth as witnessed before in the case of internationalization (and Anglification) of the national research outputs (e.g., Scandinavia and the Netherlands during the 1980s; Italy and Spain during the 1990s), a reservoir of Chinese scientists who hitherto had no access to other than national journals was tapped and now competing for access to the international literature.

China has also a large number of national journals. Zhou (2009) estimated that China had 9,468 journals in 2006, among which 4,758 in science and technology and 2,339 in the social sciences (Jiang, 2007; Ren, 2007; cf. Ren & Rousseau, 2002). China has also two citation indices in science and technology (Wu, 2004) and two more in the social sciences (Zhou *et al.*, 2010). The number of journals covered by these databases has increased during the last two decades. Thus, it seems that the growth internationally adds to China's national publication system (Jin & Leydesdorff, 2006).

Is there a justified concern about "the West loosing ground" in the sciences (Leydesdorff & Wagner, 2009a; Shelton, 2010; Wagner & Wong, 2011)? Reflexively, the bibliometric analyst can ask how reliable one can provide policy advice in these matters (Leydesdorff, 2008)? How can the bibliometric analysis be improved (Rafols *et al.*, 2011)?

## 3. Methods and Materials

All searches were performed between September 23 and 25, 2011 (unless specified otherwise), using the Web interfaces of *Scopus* and the WoS-v5 (that is, at http://www.scopus.com and http://apps.webofknowledge.com, respectively). Searches were limited to the so-called citable items: articles, proceedings papers, and reviews. Using these databases, internationally co-authored papers are attributed to contributing nations as whole numbers (so-called "integer counting"; cf. Andersen *et al.*, 1988; National Science Board, 2010). For the European Union-27, a search string with the names of all member states was composed with a Boolean OR. In the WoS, one additionally has to use "England OR Scotland OR Wales OR Northern Ireland" for the UK.

In the WoS, the years were delimited in terms of tape-years, that is, from January 1 to December 31 of each year, respectively. In *Scopus*, the corresponding search string for the USA in 2010, for example, can be formulated as follows: "AFFILCOUNTRY(United States) AND (DOCTYPE(ar) OR DOCTYPE(re) OR DOCTYPE(cp)) AND PUBYEAR is 2010". This search provides us with a replication of the report of the team of the Royal Society and Elsevier/*Scopus* (Moed *et al.*, 2011). However, the data from this search in *Scopus* includes also the social sciences and the humanities while this database enables us to exclude these domains by adding to the searches "AND NOT (SUBJAREA(Arts) OR SUBJAREA(Soci) OR SUBJAREA(Econ) OR SUBJAREA(Psyc) OR SUBJAREA(Deci) OR SUBJAREA(Busi)".[3]

The data gathering is otherwise straightforward. I distinguish additionally the group of 12 countries that joined the EU in May 2004 because these results may help to explain some of the differences between the USA and the EU-27 during the 2000s (Leydesdorff, 2000). The analysis is confined to the years 2000-2010. For the extrapolation, SPSS v.18 is used which enables users to draw the confidence intervals in the graphs.[4] The other figures are drawn from the database in Excel.

---

[3] Whereas it is not possible to search online within the *Scopus* database with the four major categories of journals—Life Sciences (> 4,300 titles), Health Sciences (> 6,800 titles, 100% Medline coverage), Physical Sciences (> 7,200 titles) and Social Sciences & Humanities (> 5,300 titles)—the website at http://help.scopus.com/robo/projects/schelp/h_subject_categories.htm offers a concordance table to 27 subject area classifications that can be searched using the function "subjarea()" in the advanced search engine of *Scopus*.

[4] The graphs are produced by the subroutine Chart Builder within SPSS. Different regression lines and curve fits can be added to the graphs in the Chart Editor. Various form of regression analysis were also performed in SPSS; for example, for determining the β coefficient.

## 4. Results

*4.1 WoS data*

Contrary to previous analyses that included also the 1990s (e.g., Jin & Rousseau, 2004; Moed, 2002), the focus on the last ten years shows that the growth of China's percentage share of publications has been increasing linearly over the last ten years ($R^2 > 0.99$). The exponential growth of China in these terms during the second half of the 1990s and the early 2000s was spectacular. Using WoS data, Figure 2 shows an extrapolation of the linear regression lines for China, the USA, and the EU-27. The decline of the EU-27 and the USA in terms of percentages of world share is partly a function of the increase of other countries (although the percentages do not have to add up to 100% given that international collaborations are counted for all contributing nations; cf. Anderson *et al.*, 1988).

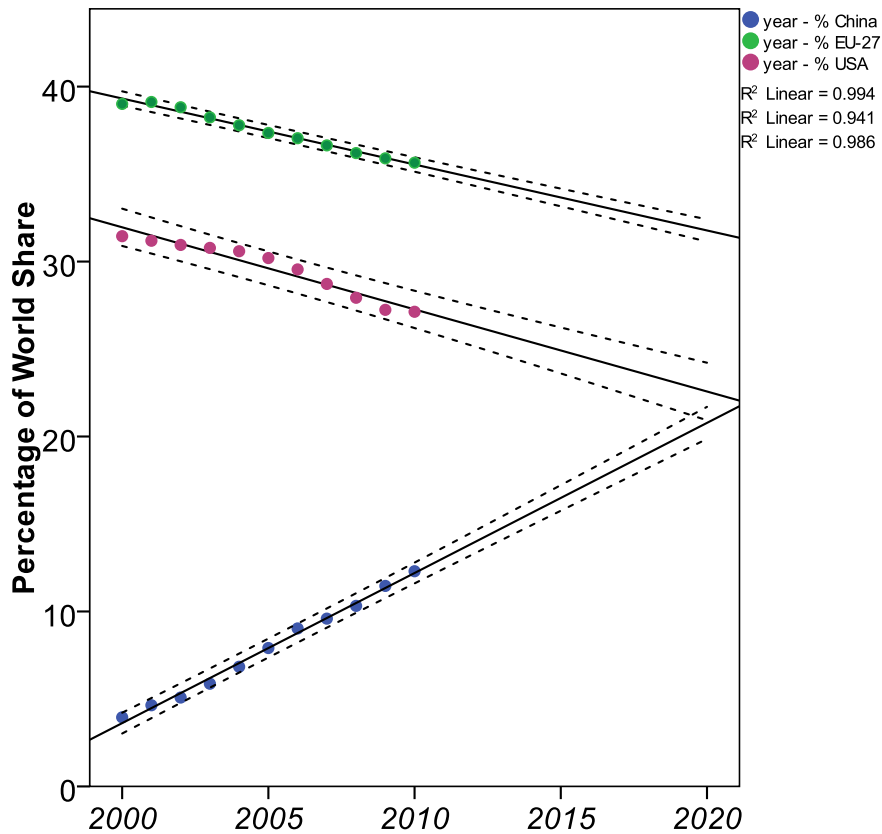

**Figure 2**: Percentages of World Share of Publications in SCIE (Articles, Proceedings papers, and Reviews) for the USA, EU-27, and China. Source: Web of Science; confidence levels indicated at the 95% level.

As against earlier predictions (e.g., Shelton and Foland, 2009; Leydesdorff & Wagner, 2009b) that found exponential growth for China (during the 1990s), the revision to linear growth in this projection leads to postponing the cross-over between the USA and China until well into the next decade. This graph predicates an even later date than a previous prediction based on using WoS.v4 data (Shelton & Leydesdorff, in press). As said, the construction of datapoints was hitherto less straightforward and perhaps less reliable (Arencibia-Jorge *et al.*, 2009).

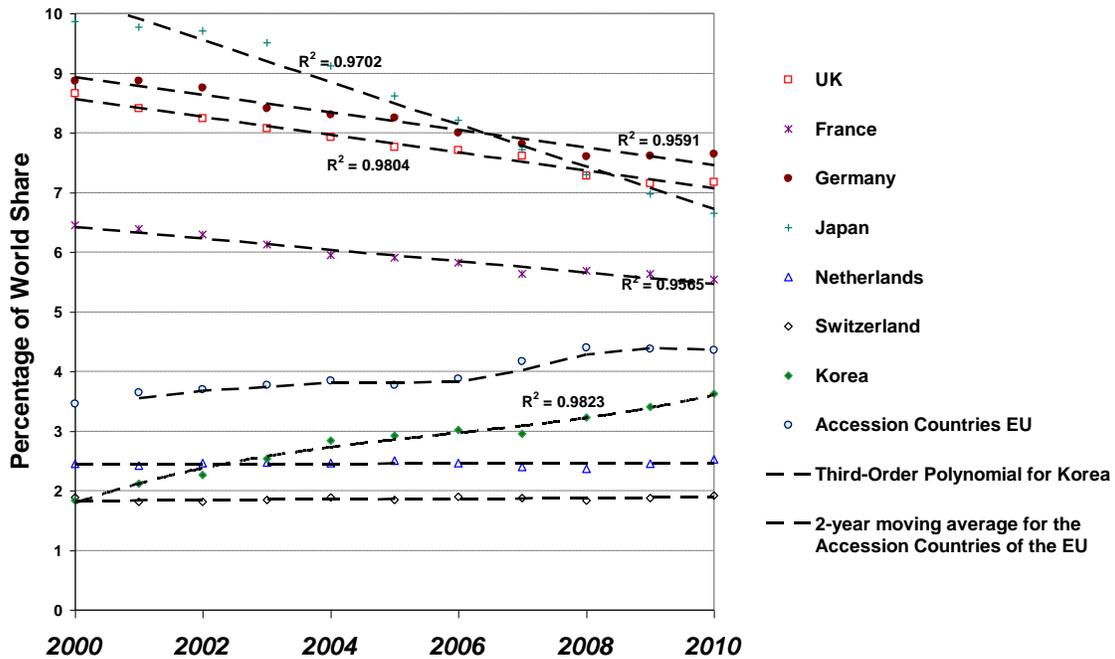

**Figure 3**: Percentages of World Share of Publications in SCIE (Articles, Proceedings papers, and Reviews) for some middle-sized and smaller nations.

Figure 3 extends the analysis to some middle-sized and smaller economies. At the top of the figure, one can see that the middle-sized countries (UK, Germany, Japan, and France; cf. Daraio & Moed, 2011) are in decline at approximately the same rate as the USA, but Japan has a steeper decline rate in the share of publications. China surpassed the UK (in this database) in 2005, and Germany and Japan in 2006.

In the lower half of Figure 2, one can see that Korea has been growing similarly to China, but this curve is not linear (Leydesdorff & Zhou, 2005; Park *et al.*, 2005; Park & Leydesdorff, 2010). The curve for Korea happens to be an excellent match for a third-order polynomial ($r^2 > 0.98$) indicating a slowing down of growth in the middle years of the decade under study. Over this whole period, the 12 new accession countries to the EU increased their shares of publication (cf. Leydesdorff & Wagner, 2009), but this growth potential seems to approach saturation during the last three years. Smaller European countries such as Switzerland and the Netherlands have been able to maintain their percentage shares during this decade; at 1.86 ($\pm$ 0.01)% and 2.45 ($\pm$ 0.01)%, respectively. This precision provides further confidence in this data.

*4.2. Scopus data*

Using *Scopus* data, one obtains a very different perspective on the shares of publications of the US, China, and EU-27 (Figure 4). The data for China again fits best with a linear regression line ($R^2 > 0.97$), but the lines for the EU-27 and the USA are shaped differently. The two or three most recent years show an upward trend that cannot be found using the WoS data. Using *Scopus*, however, the years 2010 and 2011 already fall within the 95%-confidence interval for the prediction that China might take over the first position from the USA. Thus, this effect is even stronger than the one reported previously by Clarke *et al*. (2011) and Plume (2011). However, the quality of the correlation with the linear regression is so low for the USA and the EU-27 that one can be hesitant to draw these regression lines. The confidence lines show the uncertainty.

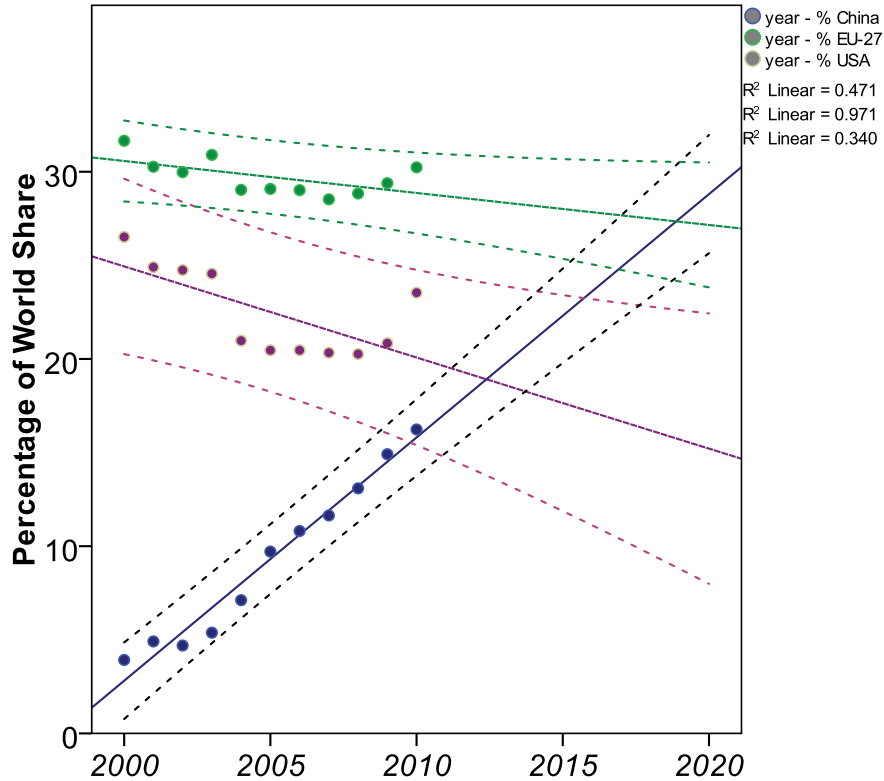

**Figure 4**: Percentages of World Share of Publications (Articles, Proceedings papers, and Reviews) for the USA, EU-27, and China. Source: *Scopus*; confidence levels indicated at the 95% level.

**5. The Social Sciences and Humanities in the *Scopus* data**

National performance using WoS data is often measured using the *Science Citation Index-Expanded* (6,650 journals) or the *Science Citation Index* (CD-Rom version; appr. 3,700 journals; National Science Board, 2010). However, the study of the Royal Society and Elsevier was based on the entire *Scopus* database including also the social sciences and humanities, while these fields are separately indexed in the WoS.

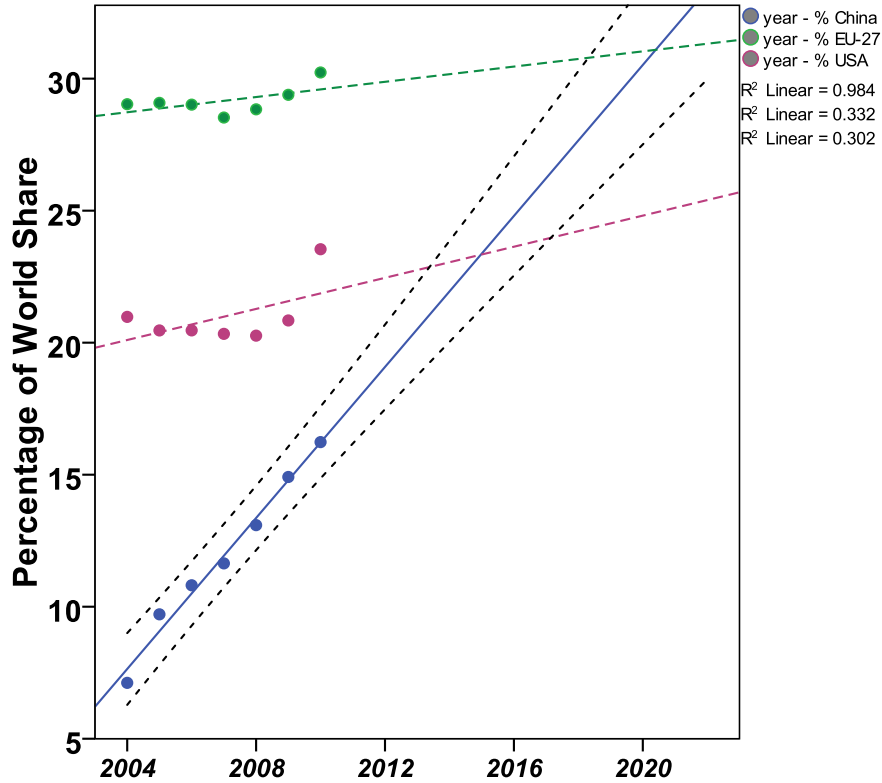

**Figure 5**: Percentages of World Share of Publications (Articles, Proceedings papers, and Reviews) for the USA, EU-27, and China, after correction for the Social Sciences and the Humanities. Source: *Scopus*, November 29, 2011; confidence levels indicated at the 95% level.

The social sciences and humanities can be excluded in *Scopus* by using the appropriate subject areas in the search string as specified above (in the methods section). The general pattern (Figure 4) does not change by this refinement, but the upward trend in the data for the EU-27 and the USA since 2004 is more pronounced than before, and highlighted in Figure 5. The message of the Royal Society/Elsevier team would be mistaken on the basis of this extrapolation of *Scopus* data.

**6. A recent revision of the prediction in *Research Trends***

In reaction to a preprint version of this paper, the staff of *Scopus* published a reply in Elsevier's online journal *Research Trends* (Moed *et al*., 2011) in which one argues that Elsevier publishes a version of *Scopus* at the internet, but also maintains a bibliometric version of this database in which the data is subjected to more intensive cleaning processes. As stated: "Especially the results for the USA differ considerably between the two *Scopus* versions. These discrepancies are due to the fact that not all author affiliations contain the name of the country in which the author's institutions are located. This is especially true for US affiliations: many indicate the US State but not the country name."

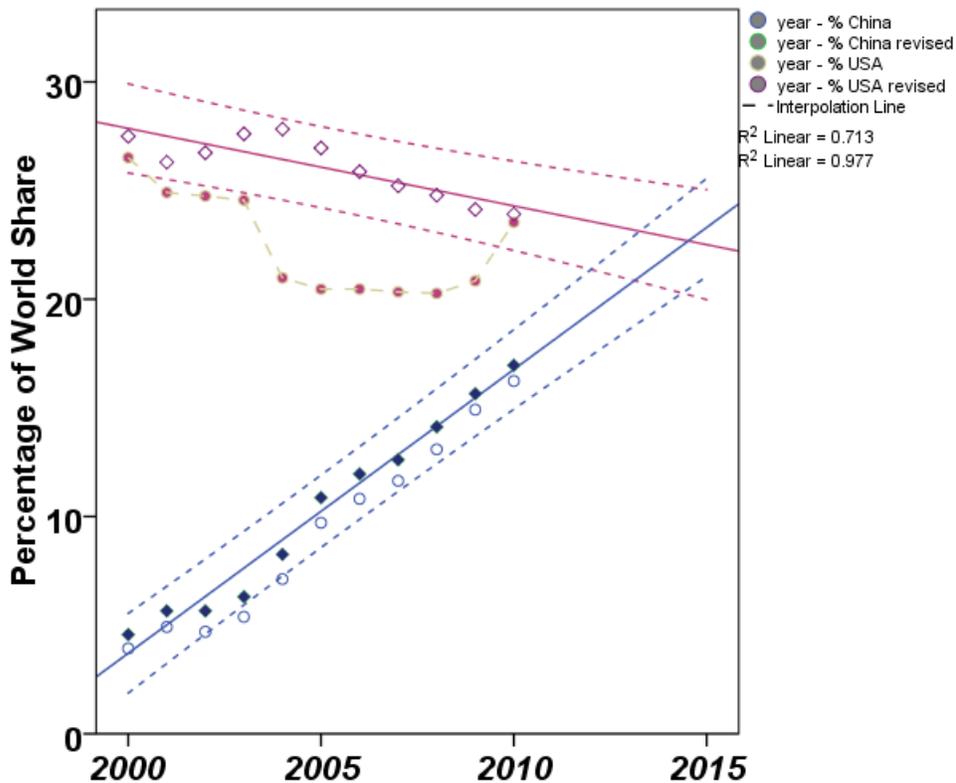

**Figure 6**: Revised Percentages of World Share of Publications (Articles) for the USA (◊) and China (♦). Source: the bibliometric version of *Scopus*; confidence levels indicated at the 95% level.

Figure 6 is based on a reconstruction of this revised data of the *Scopus* team (Leydesdorff, 2011; Moed *et al.*, 2011). The values from Figure 4 are penciled in for the comparison. Indeed, the differences are largest for the USA in almost all years. However, even in this corrected data the previous prediction of a cross-over in 2013 is not fully warranted and the fit for the linear regression in the case of USA data remains relatively poor ($R^2 = 0.71$).

**7. Discussion**

What might cause these differences between the measurements in the respective databases? Let me first stipulate that in both databases I used 2000-2010, whereas the team of the Royal Society and Elsevier used 1996-2008 for their prediction. When this report appeared in March 2011, I replicated the measurement and found some deviation for points for 2009 and 2010, but assumed that this could be an artifact because the publication year 2010 was not yet completed by March/April 2011. Publications may arrive with the time-stamp of 2010 at a later date in 2011, and practice may vary for publications from different world regions. However, a repeat of the measurement in September did not change these results.

I deliberately used the data since 2000 because *Scopus* data are only reliable since 1996 (Ove Kahler, *personal communication*, 28 August 2009), and the database was gradually improved in terms of coverage during the initial years. As against the Web of Science, *Scopus* claims to include more regional journals among the 18,000 journals covered by this database (see at http://www.info.sciverse.com/*Scopus*/*Scopus*-in-detail/facts).

The *Web of Science* nowadays covers approximately 11,500 journals including approximately 3,000 journals added since 2008 (Testa, 2011). Thomson Reuters first announced this as an expansion of regional coverage in May 2008, possibly in response to competition from *Scopus*. The comparison of Figures 2 and 4 above, however, teaches us that the focus in the *Web of Science* has remained on Europe and the USA more than in *Scopus*.

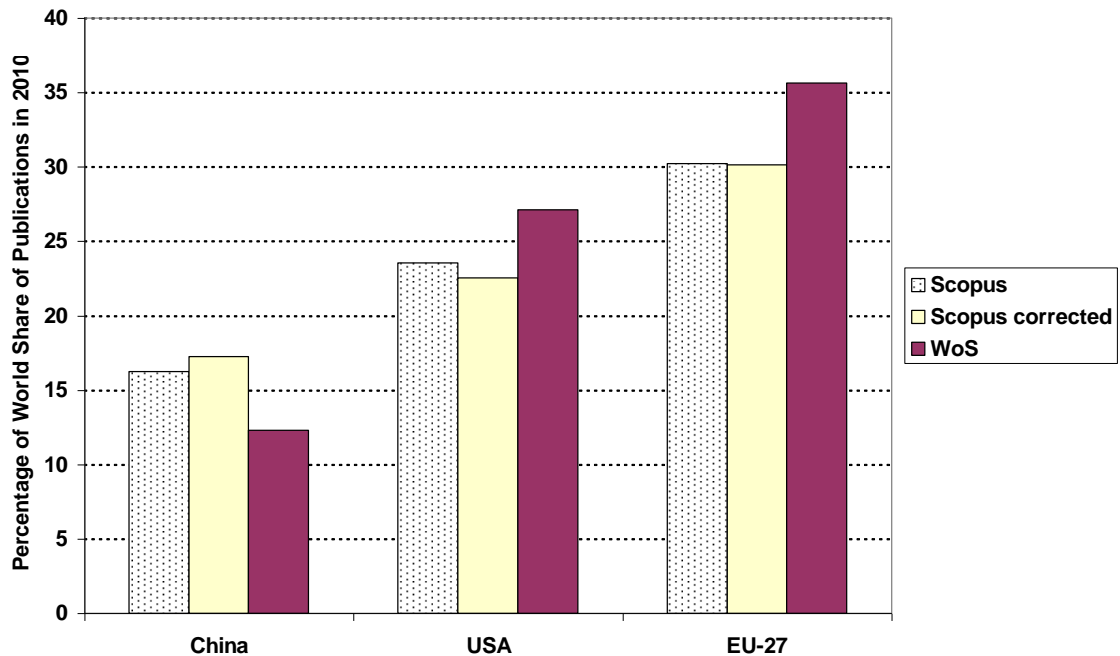

**Figure 6**: Percentages of World Share of Publications for China, the USA, and the EU-27 in 2010 in both *Scopus* (with and without correction for the Social Sciences and Humanities) and WoS (v5).

Figure 6 shows that the percentage of share of Chinese publications in the WoS (v5) is 12.30% in the Web of Science, while it is 17.24% in *Scopus* (after correction for the social sciences and humanities). Similarly, the USA has 22.54% in *Scopus* data as against 27.13% in the Web of Science. The differences are approximately five percentage points on either side, and thus add up to more than 9.5%. For the EU-27, the difference between the two databases is even larger, with 30.12% in *Scopus* and 35.65% in the WoS or a difference of 5.53 percentage points; this difference is of the size of the contribution of France.

## 8. Conclusions

The bibliometric contribution to the policy debate about the ranking of national and institutional science systems, in my opinion, should focus on the specification of uncertainty and possible sources of error and potential misinterpretation (cf. Leydesdorff, 2008). In a seminal paper, Martin & Irvine (1983) suggested to rely on "converging partial indicators" for the assessment: results of the bibliometric analysis can be considered as more reliable when the results indicate the same trends or differences in rankings. Using the *Scopus* database, however, one could even make a case for a relative increase of the shares of publications for the US and the EU since 2004 (on the basis of Figures 4 and 5 above).

The confidence lines and the fits provide in the above comparisons between WoS and Scopus data an argument to build for policy advice preferentially on WoS data since the uncertainty is lower. However, the difference in coverage between these two major databases is significant at this high level of aggregation: *Scopus* is more oriented to the Chinese publication system and less to the US and the EU than *SCIE.* As noted, the differences can be in the order of five to ten points. In my opinion, such differences are worrysome and worth to be noted in the case of policy advice (Clarke et al., 2011; Rafols *et al.*, in press).

Although strong growth remains indicated for the case of China, the USA cannot be expected to continue declining linearly. Whereas the world sum of publication is not a zero-sum game because of the steady increase of international coauthorship relations (Persson *et al*., 2004; Wagner, 2008), the competition drives in the direction of decreasing marginal returns because all nations are investing in order to improve their share of publications (and citations). In addition to the dynamics of the competition, the above exercise reminds us that the dynamics of the databases also need to be taken into account.

**Acknowledgement**
I am grateful for comments of Lutz Bornmann and Henk Moed.